# First investigation of the response of solar cells to heavy ions above 1 AMeV

A. Henriques[a,*], B. Jurado[a,†], J. Pibernat[a], J. C. Thomas[b], D. Denis-Petit[a,c], T. Chiron[a], L. Gaudefroy[c], J. Glorius[d], Yu. A. Litvinov[d], L. Mathieu[a], V. Méot[c], R. Pérez-Sánchez[a,c], O. Roig[c], U. Spillmann[d], B. Thomas[a], B. A. Thomas[a], I. Tsekhanovich[a], L. Varga[d], Y. Xing[d]

[a] Centre d'Etudes Nucléaires de Bordeaux Gradignan (CENBG), CNRS/IN2P3, Université de Bordeaux, 19 Chemin du Solarium, 33170 Gradignan, France
[b] Grand Accélérateur National d'Ions Lourds (GANIL), Bd Henri Becquerel, BP 55027-14076 Caen, France
[c] CEA, DAM, DIF, F-91297 Arpajon, France
[d] GSI Helmholtzzentrum für Schwerionenforschung, 64291 Darmstadt, Germany

**Abstract**

Solar cells have been used since several decades for the detection of fission fragments at about 1 AMeV. The advantages of solar cells regarding their cost (few euros) and radiation damage resistance make them an interesting candidate for heavy ion detection and an appealing alternative to silicon detectors. A first exploratory measurement of the response of solar cells to heavy ions at energies above 1 AMeV has been performed at the GANIL facility, Caen, France. Such measurements were performed with $^{84}$Kr and $^{129}$Xe beams ranging from 7 to 13 AMeV. The energy and time response of several types of solar cells were studied. The best performance was observed for cells of 10x10 mm$^2$, with an energy and time resolution of $\sigma(E)/E=1.4\%$ and 3.6 ns (FWHM), respectively. Irradiations at rates from a few hundred to 10$^6$ particles per second were also performed to investigate the behavior of the cells with increasing intensity.

## 1. Introduction

Solar cells, the devices used to convert the energy of sunlight into electricity, appear as a very interesting and cost-efficient option to detect heavy ions. Solar cells, also referred as photovoltaic cells, were first used in 1979 by Siegert to detect fission fragments produced by the interaction of thermal neutrons with actinide nuclei [1]. The produced fission fragments cover a broad range of nuclei ranging from mass number A=60 to 160 with a typical kinetic energy of 1 AMeV. At the time, several advantages were already identified, such as the low cost, flexible geometry and the quality of the response to fission fragments (a FWHM energy resolution of 1% to 2% was reported).

Some years later, Ajitanand et al. highlighted the solar cells radiation hardness as well as their capability to detect fission fragments in an intense background of light charged nuclei [2]. In 1987, Liatard et al. exposed solar cells to scattered ions of $^{12}$C up to energies of 240 MeV revealing a linear energy response just up to 80 MeV [3]. This study also measured the time resolution between two cells of 10 mm$^2$ to be 12 ns FWHM and pointed out the dependence of the time response on the cell size. Since then, solar cells have been used in a few experimental campaigns as fission fragment detectors, see e.g. [4-7]. They have often been used as heavy ion counters and to perform coincidence measurements [7].

[*] Present address: henrique@frib.msu.edu, FRIB/NSCL Michigan State University, 640 South Shaw Lane, East Lansing, MI 48824, USA
[†] Corresponding author : jurado@cenbg.in2p3.fr

*1.1. Charge collection*

Similar to silicon detectors, solar cells present a semiconductor structure but with a smaller depletion zone, usually about 1 µm. As a consequence, the capacitance of solar cells is around 40 nF/cm$^2$, typically a thousand times larger than the capacitance of silicon detectors. Contrary to Si detectors, no bias voltage is required, in fact the application of a bias voltage increases the noise level due to an increase of the reverse current [1].

The origin of the electric field can be intriguing given that solar cells do not require any bias voltage. The importance of the built-in voltage of solar cells in charge collection is mentioned in [8] for the collection of photo-generated charges. Also referred is the role of the widths of the different layers, charge mobility and lifetime on the enhancement of charge collection by drift.

In the usual mode of operation of a solar cell, when a photon strikes a cell, electron-hole pairs are created at the junction and the charge is collected mainly by diffusion. However, when a charged particle impinges on a solar cell, the effect is different. In 1981, Hsieh explained the severe transient distortion that takes place in the depletion zone when an alpha particle impinges on a silicon device and its role in the charge collection, the so-called field-funneling effect [9].

After the passage of a charged particle, an electron-hole plasma column is created along the particle track. This plasma density is usually orders of magnitude greater than the substrate doping density, neutralizing the initial junction depletion zone that is close to the track. In addition, the electrons that are directed towards the positive electrode cancel the electric fields of the junction. The plasma drives the electric fields into the substrate, along the particle track. The plasma column tends to spread radially and this enables the separation of electron and holes. It allows the charge collection to occur by drift and diffusion, in opposition to just diffusion as in the normal mode of operation. The electrons drift along the plasma column and are collected by the electrode. As the plasma density reduces, the depletion layer begins to reform until it is completely regenerated. The funneling efficiency is a strong function of the energy loss profile dE/dx, leading to very weak signals for light charged particles of few MeV, which cannot be detected.

It is not our aim to deeply describe the complex funneling effect, a qualitatively description can be found in [10] and its model in [11]. However, its application to solar cells is yet to be described. The lack of predictive power supports the need for measurements considering heavy ions with energies above 1 AMeV.

*1.2. Motivation*

Solar cells are an alluring option for the detection of heavy ions at radioactive beam facilities, e.g. to detect heavy reaction products in nuclear physics experiments. They can also have a great impact as beam intensity monitoring devices.

Moreover, the cells radiation hardness positions them as a very interesting option to be used in challenging and stringent vacuum environments like inside heavy-ion storage rings. Indeed, replacing damaged detectors implies venting the ring and re-establishing ultra-high vacuum (UHV), $10^{-10}$ to $10^{-12}$ mbar, even in a small part of a ring, takes several days. The outgassing rate of solar cells and detector supports will be investigated at the CENBG. A preliminary measurement showed a very low outgassing, bellow $10^{-11}$ mbar·l/(s·cm$^2$) after baking for 48 hours at 200°C. The cells were operational after baking. This and the possibility to use them as counters in coincidence measurements, makes solar cells an interesting option to be considered in our future measurements at storage rings [12].

The present work is a first exploratory measurement of the response of solar cells to heavy ions up to energies of 13 AMeV.

The energy and time response and the behavior with increasing beam intensity of solar cells were studied at the GANIL facility in Caen, France. $^{84}$Kr and $^{129}$Xe beams were used at energies from 7 to 13 AMeV. Two types of cells of different sizes were investigated, representing two types of composition and structure. The investigated cells were monocrystalline used in different applications: for energy production on Earth (roofs cells) [13] and in space (space cells) [14]. Their sizes ranged from 10x10 mm$^2$ to 30x30 mm$^2$ and the thickness between 220 and 250 μm. The composition and structure of the cells can have implications in the formation of the electrical signal and therefore in the response of the solar cell when exposed to heavy ions. As shown in section 3, the size of the cell has implications in the cell capacitance influencing the output signal and the electrical circuit that will follow, in particular the pre-amplifier.

## 2. Characteristics of solar cells

The composition of the solar cells was determined via the Rutherford Backscattering method using alpha particles of 2 MeV at the AIFIRA facility [15] in Bordeaux, France. This analysis allowed us to identify the main components of each cell given their different architecture. But mainly, it was verified that the roof cells had a substrate of silicon while the substrate of the cells used in space applications was germanium and for both the active layers were around 1.2 μm. Figure 1 shows a schematic drawing of a solar cell and Table 1 lists the solar cells we used during the experiments.

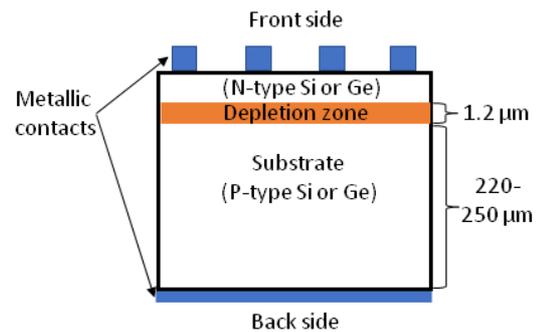

*Figure 1 – Simplified diagram of the structure of the used solar cells.*

From the difference in materials in the substrate, one can expect different output signal amplitudes as the energy needed to create an electron-hole pair in silicon and germanium are 3.6 eV and 2.9 eV, respectively. For ideal detectors, if an ion of $^{129}$Xe impinges at 10 AMeV, the total collected charge for the silicon detector is 57 pC and 71 pC for the germanium detector. Assuming the same collection efficiency for both types of cells, this will affect the rise time and the amplitude of the output signal.

| Nomenclature | Supplier | Application | Substrate | Size (mm$^2$) | Number of cells |
|---|---|---|---|---|---|
| 10x10S | Solar Made | Household panels | Silicon | 10x10 | 3 |
| 10x10G | SpaceAzur | Space | Germanium | 10x10 | 2 |
| 20x20S | Solar Made | Household panels | Silicon | 20x20 | 2 |
| 30x30S | Solar Made | Household panels | Silicon | 30x30 | 1 |

Table 1- List of the solar cells used during the experiments: used name, product supplier, main application, main element of the substrate, size and number of cells tested during the measurements.

The germanium-substrate cells were cut using a silver wire, whereas the silicon-substrate cells were purchased with the correct sizes. The cells were cleaned through an ultra-sound bath using ethanol (96%) before the measurements. The response of cells of the same type and size was within the observed uncertainties. This was verified by comparing the signals induced by fission fragments originating from a $^{252}$Cf source and during irradiation with the higher-energy ions at GANIL. We found that the reproducibility of the results was dominated by the quality of the electric contact.

## 3. Electronic model of solar cells

The solar cell electronic model is presented in Fig. 2. The model was verified by performing impedance measurements using the Potencio-Electrochemical Impedance Spectroscopy technique at the IMS laboratory of the University of Bordeaux, considering silicon cells of different sizes.

The model consists on one capacitor (Cd) in parallel with a resistor (Rp), which are in their turn connected in series with the resistor Rs. The current (id) generated by an impinging particle is represented in the circuit with a current generator. The values of Rp (5 kΩ), Rs (0 -10 Ω) and Cd (38 nF/cm$^2$) were determined. Such information allowed us to obtain the transfer function of the electronic circuit shown in Fig. 2:

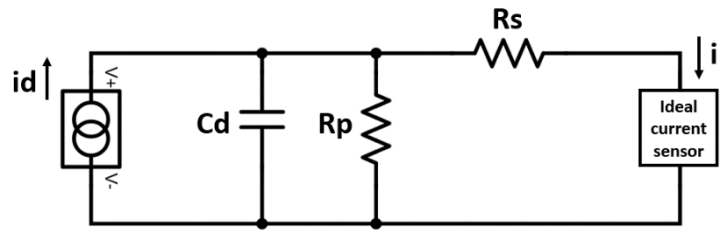

Figure 2 - Solar cell electronic model considered for equation 1.

$$i = id \frac{1}{1 + j\frac{Rp \cdot Rs}{Rp + Rs} Cd\, \omega}$$

Where $i$ is the amplified output current. When considering that $Rs \ll Rp$, the transfer function can be simplified to:

$$i = id \frac{1}{1 + jRsCd\, \omega} \quad (1)$$

The latter expression reveals a low pass filter behavior with a cutoff frequency ($f_c = \frac{1}{2\pi RsCd}$) dependent on Cd and Rs values. In the frequency domain, a large capacitance translates in a lower cutoff frequency. While in the time domain, the integration or time constant (τ=RC) is larger and therefore for the same pulse duration, one obtains smaller amplitudes for larger capacitances and thus, larger solar cells.

## 4. Experimental set-up

At the GANIL facility, the CIME cyclotron was used to accelerate beams of $^{84}$Kr at 7 and 10 AMeV and of $^{129}$Xe at 10 and 13 AMeV, covering a total energy range from 588 MeV up to 1677 MeV.

The solar cells were mounted on a rotating stainless-steel support that could house up to 9 cells (Fig. 3). The rotating support was inserted into the beam line with a propulsor. With the aid of a goniometer, each cell was positioned and irradiated at a time. In one of the positions, a silicon detector was placed for a reference measurement: an ORTEC surface barrier silicon detector with an active area of 100 mm$^2$, a depletion depth of 300 μm and a guaranteed resolution of 14 keV for 5 MeV alpha particles. The bias voltage of this detector was 40 V.

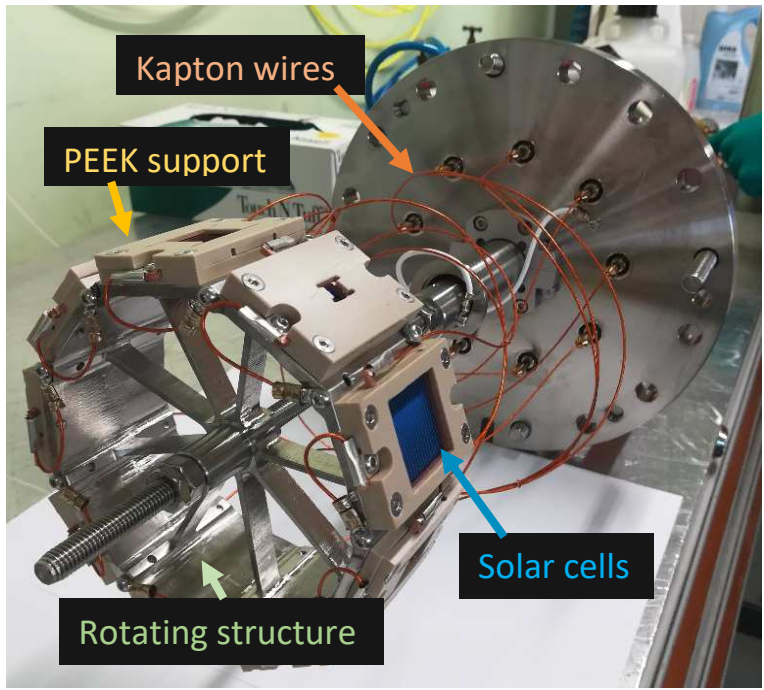

Figure 3- Rotating holder with the mounted solar cells from Table 1, exposed to heavy ion beams at GANIL.

The rotating stainless-steel structure was set as the electrical ground. It was placed at the exit of the CIME cyclotron. A gas profiler allowed us to evaluate the beam spot size, which had a typical size of 5 mm and 7 mm in X and Y direction, respectively. Each cell at its turn was placed perpendicular to the beam.

The solar cells were sandwidched between two rectangular shaped pieces of PEEK (PolyEtherEther Ketone). Copper frames, integrated on each PEEK structure, allowed the collection of the signal from the front part of the cell and the connection to the ground set at the backside of the cell.

Since we foresee to use solar cells in UHV conditions, only mechanical contacts were used to support the cells, avoiding any welding. Kapton insulated cables were used to profit from their excellent electrical insulation properties and low outgassing rates. The connection of the signal cables with the single ended BNC feedthroughs on the flange was done using BeCu connectors.

Regarding the electronic set-up, the used pre-amplifier device, consisting of a transimpedance pre-amplifier (i.e. a current to voltage converter) and a fast shaper, was the one previously developed for experimental campaigns aiming at detection of low-energy fission fragments (1 AMeV) [16]. An example of the output signal of the pre-amplifier device, for a 10x10 mm$^2$ silicon cell exposed to $^{84}$Kr beam at 7 AMeV, can be seen in Fig. 4.

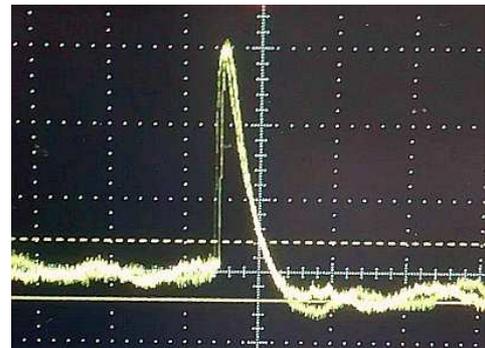

Figure 4 - Signal at the output of the pre-amplifier of a 10x10 mm$^2$ silicon cell from $^{84}$Kr beam at 7 AMeV. The scale was set to 200 mV/division and 1 µs/division vertical and horizontally, respectively.

The RMS noise level for all the solar cells sizes was around 6 mV. The noise level is the same because the low-pass filter behavior of the solar cells ensures that the high frequency noise component is removed for all cell sizes. Nevertheless, the signal to noise ratio was dependent on the solar cell size, as expected from eq. 1.

Fig. 5 presents the scheme of the electronic chain used. The signal after the pre-amplifier device (Fig. 4) was delivered to a linear amplifier and a fast amplifier. Through a constant fraction discriminator (CFD), the output signal from the fast amplifier generated the trigger signal, opening a gate. The gate defined the time during which the peak-sensing

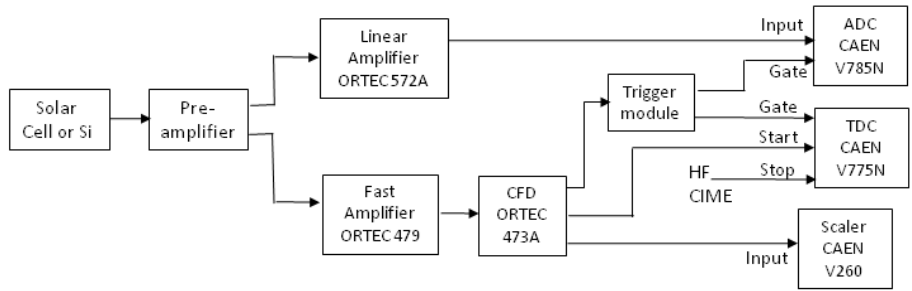

*Figure 5 – Scheme of the electronics used in the present measurements.*

Analog to Digital Converter (ADC) would track the maximum of the output signal of the amplifier. Also, the output of the CFD was used as a START signal in a Time to Digital Converter (TDC), while the STOP signal was provided by the operating frequency (HF) of the CIME cyclotron (typically in the 10 MHz range). The output of the CFD was also sent to a Scaler to measure the frequency and number of events. Except for the trigger module, all the used electronic modules are commercially available. The names of the modules are given in Fig. 5.

## 5. Results

*5.1. Signal features at the output of the pre-amplifier device: rise time, fall time and amplitude*

The characteristics of the output signals (rise time, fall time and amplitude) of the pre-amplifier device were observed with the help of an oscilloscope for each cell and each beam.

The rise time ranged from 200 to 1300 ns as shown in Fig. 6. The rise time was evaluated for the different substrates (silicon (S) or germanium (G)) (Fig. 6-left) and the different cells sizes (Fig. 6-right). For each substrate, a dependence on the energy is observed: higher beam energies are associated to higher rise times. In addition, the rise time is always lower for cells whose main substrate element is germanium, which can be explained either by the signal collection process or by the expected lower capacitance of a germanium junction [17]. Looking at the dependence with the cells size, the larger rise times were registered for larger cells. This is in agreement with the expected low pass filter behavior as larger cells have a larger capacitance. Regarding the fall time, it ranged between 1.6 to 16 μs and the dependence with the substrate main element and size was similar to the rise time.

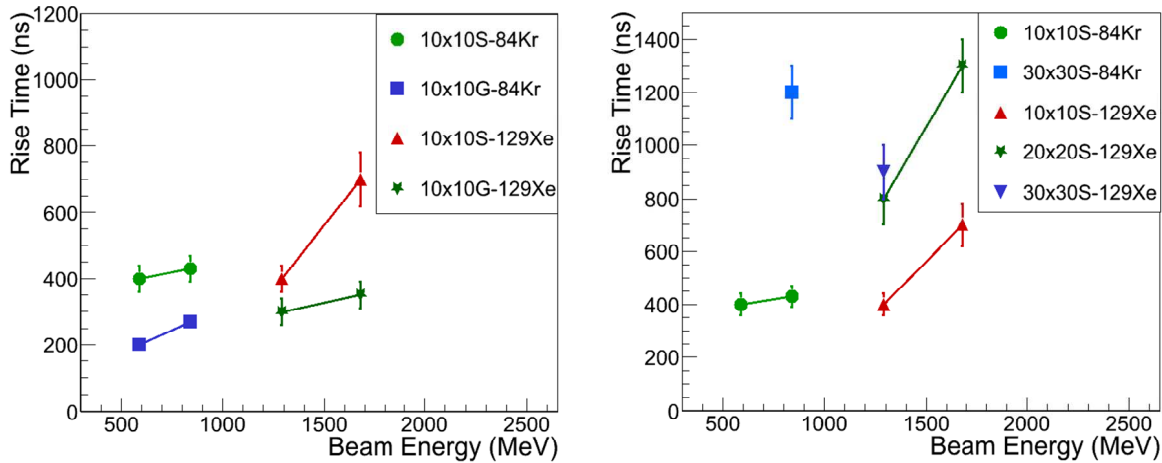

*Figure 6 - Average rise time observed for each beam energy considering: (left) the two different substrates (silicon (S) and germanium (G)) for 10x10 mm$^2$ cells and (right) different cell sizes (10x10, 20x20 and 30x30 mm$^2$). The rise time corresponds to the time needed to go from 0 to 100% of the signal amplitude.*

The average amplitude of the output signal of the pre-amplifier device was also measured, for the different beam energies and cells, and its values are presented in Fig 7.

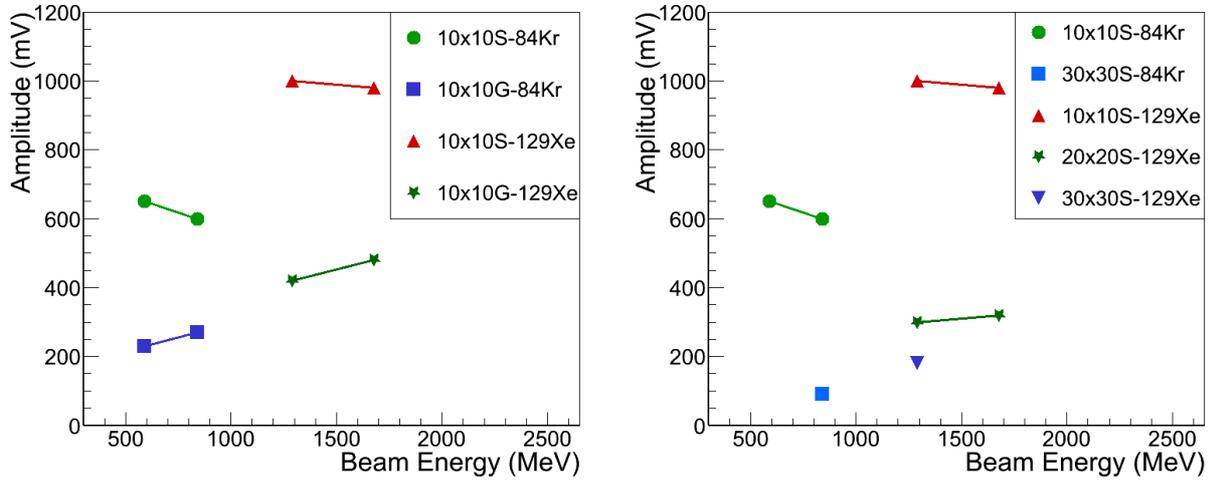

*Figure 7 - Average amplitude observed for each beam energy considering: (left) silicon (S) and germanium (G) substrates for 10x10 mm$^2$ cells and (right) the different cell sizes (10x10, 20x20 and 30x30 mm$^2$).*

A higher amplitude of the pre-amplifier output signal is observed for silicon substrate cells (Fig. 7- left). This was not the anticipated behavior due to the larger number of charges produced in the germanium substrate and the expected smaller capacitance when compared to silicon, and might be explained by a lower charge collection efficiency for the germanium type cells. On Fig. 7-right, it is shown the amplitude for different sizes of solar cells of silicon substrate. It is observed that larger cells provide a smaller signal amplitude. Such dependence with the cell size is well understood with the electronic model of solar cells, as larger cells have a larger capacitance. In fact, we observe that the ratio of the amplitudes is fairly close to the inverse of the ratio of the surfaces.

Regarding the energy dependence, we observe a general increase of the amplitude with the beam energy, except for the 10x10 mm$^2$ cell with silicon substrate, which tends to show a slight decrease. This behavior is not yet clear to us and needs further investigation.

*5.2 Energy and Time Resolution*

Spectra like the ones of Fig. 8 allowed us the characterization of solar cells in terms of energy and time resolution. The time spectrum was calibrated with a time calibrator. On the energy spectrum shown on the left part of Fig. 8, a tail is visible on the left side, which can be related to pile-up events during the fall-time of the pre-amplifier signal where an undershoot is observed (Fig. 4). In addition, for each beam energy the silicon detector was placed in line to have a point of comparison and to control the beam quality.

The energy and time resolution were obtained for the different ion beams, energies and cell types. The presented values in Fig. 9 for the energy resolution refer to the ratio of the standard deviation and the mean value of the distribution, while the time resolution was obtained via the FWHM of the distribution. These results were obtained using a Gaussian fit (as seen on Fig. 8). For such measurements the beam intensity was of a few hundred particles per second (pps).

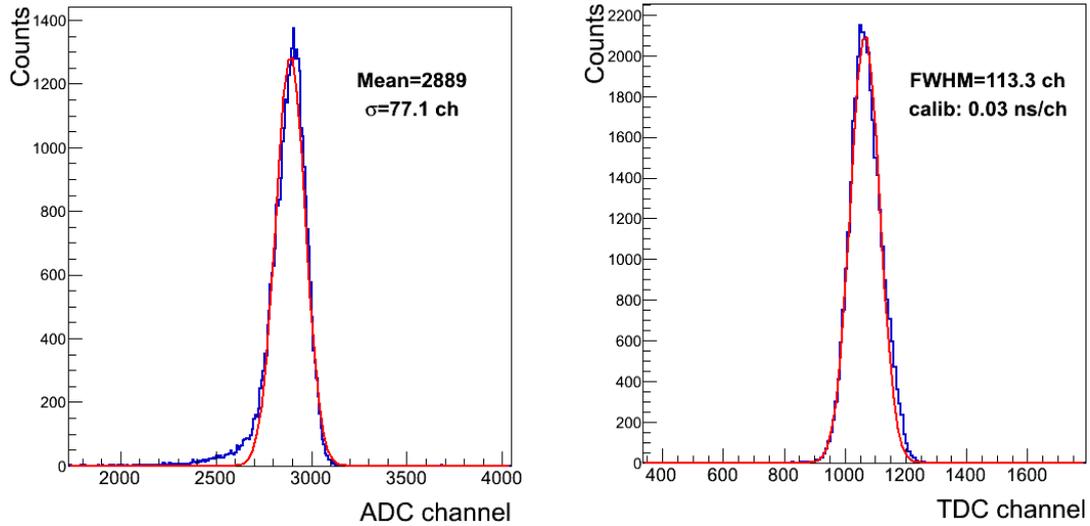

*Figure 8 – Energy (left) and time (right) spectra of a 10x10 mm² silicon cell irradiated with a $^{129}$Xe beam at 10 AMeV. The time spectrum was obtained from the time difference between the cell signals (START) and the cyclotron frequency signal (STOP). The calibration value to convert channels into ns is given. The Gaussian fits represented by the full red lines provided the standard deviation $\sigma$ for the energy and time resolutions.*

The energy resolution ranged from 1.2 and 2.9%. The best result was observed for a germanium substrate 10x10 mm² cell: 1.2%. Although the energy response of the overall beam settings was to some extent better for the germanium substrate cells, the results for the silicon substrate do not lay far from it, as observed in Fig. 9-left.

From Fig. 9, 10x10 mm² cells present in general a better energy resolution, between 1 and 3%. Once again, this can be explained by the lower capacitance of the smaller solar cells that provide a better signal-to-noise ratio response. When comparing the solar cells results with the silicon detector, the latter provides an energy resolution of 1% or better for the same beam settings. The energy resolution of the beam delivered by the CIME cyclotron is typically 0.5%. Therefore, the measured energy resolutions are dominated by the detectors response.

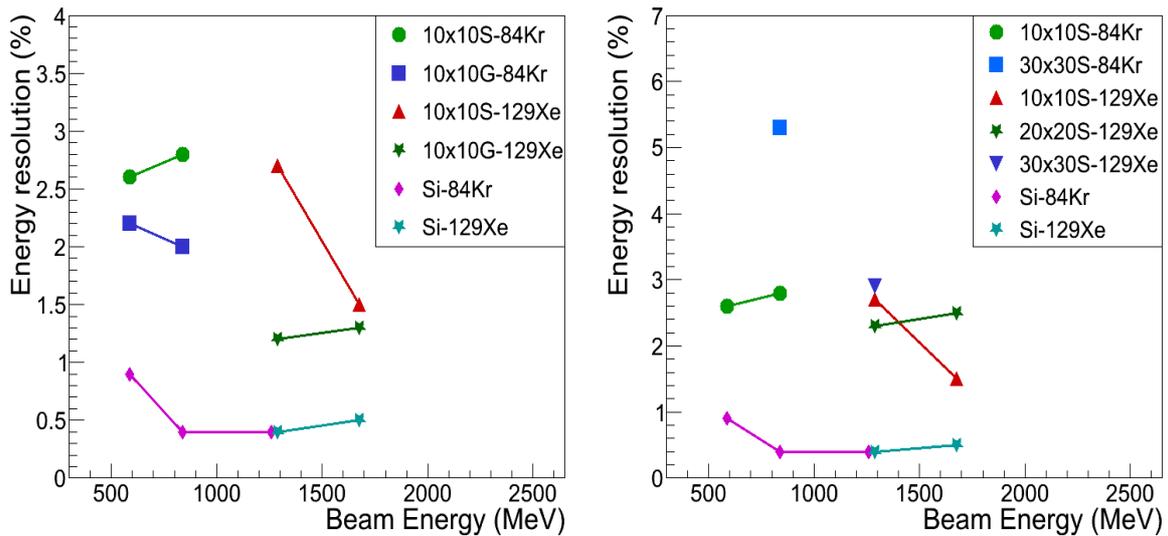

*Figure 9 - Energy resolution (standard deviation over mean value) observed for each beam energy considering (left) the two different substrates (silicon (S) and germanium (G)) for 10x10 mm² cells and (right) the different cell sizes (10x10, 20x20 and 30x30 mm²). The results obtained with the silicon detector (Si) are also shown.*

The results for the time resolution are presented in Fig. 10 for all cells sizes considered in this exploratory study. The time resolution ranged from 3.6 to 14 ns, from which the best result corresponds again to the 10x10 mm$^2$ size. Here, similarly to the energy response, it is also observed that the smaller the size, the better is the time response. As follows from Fig. 10, the time response of a silicon detector is always better than the response of the solar cells, being between about 1 and 4 ns. The main contributions to the time resolution are the detector response and the time spread of the pulses delivered by the CIME cyclotron. Assuming a time resolution for the Si detectors of about 0.8 ns, which is associated to the minimum value in Fig. 10, we deduce that the time spread of the pulses varies between 0.6 and 4 ns, depending on the beam energy. Therefore, we conclude that the time resolution of the solar cells is always dominated by the detector response.

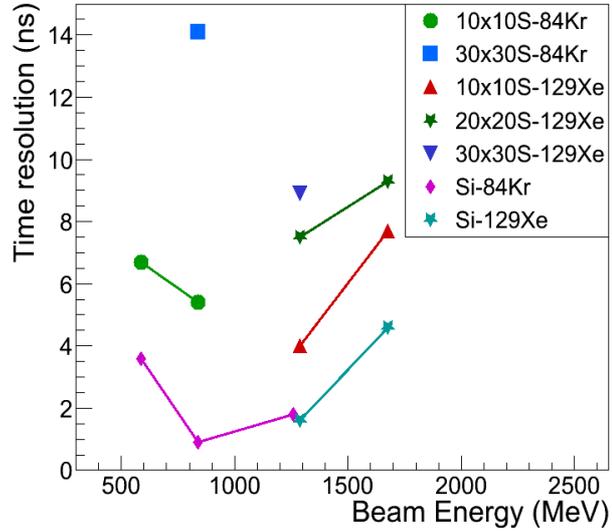

Figure 10 - Time resolution (FWHM) observed for each beam energy considering the different sizes: 10x10, 20x20 and 30x30 mm$^2$ for silicon (S) substrate cells. The results obtained with the silicon detector (Si) are also shown.

The presented measurements with the $^{84}$Kr and $^{129}$Xe beams, together with the characterization of the solar cells, allowed us to develop a new pre-amplifier for the solar cells which was successfully tested at GANIL using a $^{238}$U beam.

*5.3. Behavior as a function of the beam intensity*

Radiation damages have been investigated thoroughly for silicon detectors [18-21] and even considering heavy ions [22-24]. According to Shiraishi et al., fission fragments produce more defects than lighter particles [22]; Kurokawa et al. go further and evaluate the damages in silicon detectors as being $10^3$ to $10^5$ times larger for heavy ions than for protons [23]. When a heavy ion impinges on a silicon detector, it can create a defect that can change the energy gap level of the material. Locally, it can create an electron emission center which will be the source of a leakage current. It can also decrease the output pulse-height due to recombination of charge carriers and lower the energy resolution. The bias voltage can be increased to compensate for the incomplete charge collection.

In the case of solar cells, such detailed studies have not yet been performed. Nevertheless, the integrated flux and the pulse-height for a solar cell and a surface barrier detector were compared using fission fragments from a $^{252}$Cf source: for an integrated flux of $10^7$ fragments/cm$^2$ a solar cell loses 10-15% of its pulse-height, while the surface barrier detector loses 50% [25]. It was also reported that bombarding a solar cell with $10^9$ protons/s during 30 minutes had no effect on the performance of the solar cells in the detection of fission fragments from a $^{252}$Cf source [2].

We have conducted a first test on radiation resistance of solar cells. Several cells were irradiated at different energies and intensities. The intensity ranged from a hundred of pps up to one million pps. Most of the cells used in this test, had already been irradiated during the energy and time resolution evaluation, the aim of this study was to verify the stability of the measurements with time and different beam intensities.

These measurements were possible due to the beam intensity reduction devices available before injecting the beam into CIME. As discussed in section 4, the total number of signals delivered by the cell during irradiation was recorded by a scaler. In addition, the signal from a pulse generator with 20 Hz frequency was fed into the scaler to provide a time reference.

The evolution of the energy and time response of the cells can be observed by plotting the ADC and TDC channels against the accumulated number of registered events. Such temporal evolution is shown in Fig. 11, where a 7 AMeV $^{84}$Kr beam was used to irradiate a 10x10 mm$^2$ silicon cell. From such measurement it was verified that for a low rate of 470 pps the energy and time response do not change over 1 minute of irradiation.

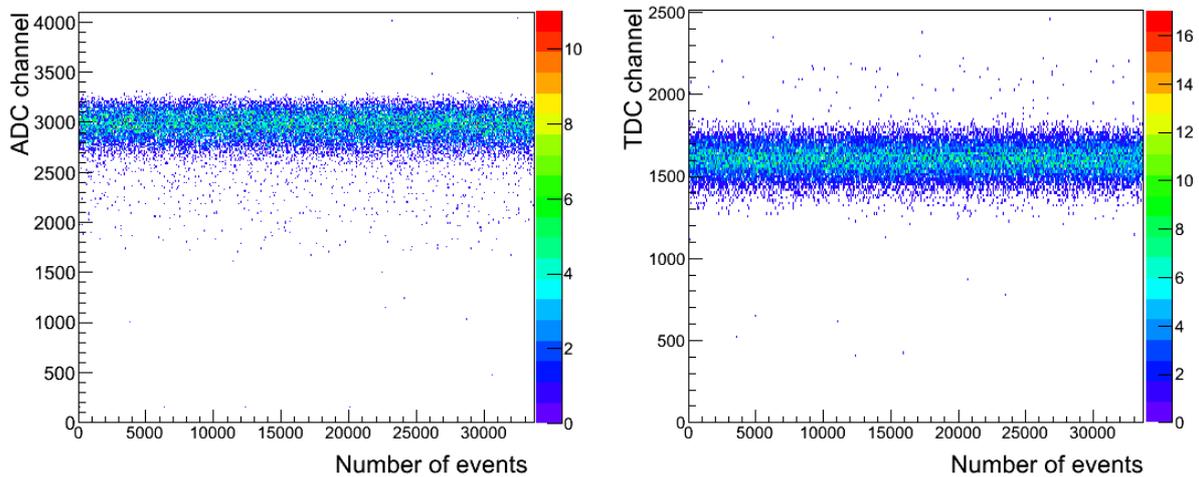

Figure 11 – Energy and time response in arbitrary units as a function of the accumulated number of events for a 10x10 mm$^2$ silicon cell irradiated during one minute at a rate of 470 pps with a 7 AMeV $^{84}$Kr beam.

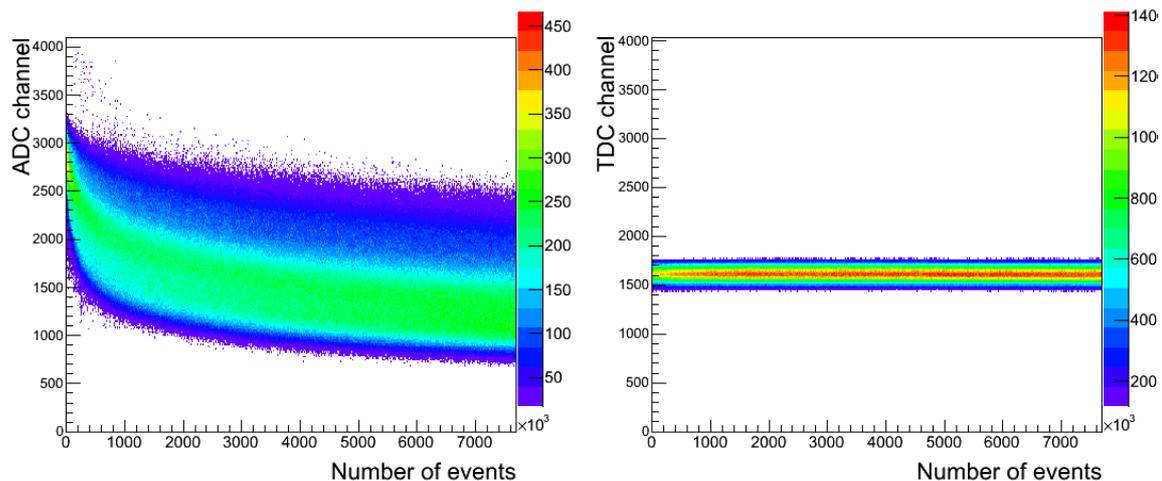

Figure 12 - Energy and time response of a 10x10 mm$^2$ cell irradiated with an $^{84}$Kr beam at 7 AMeV with a beam intensity of 53 thousand pps.

When increasing the beam intensity up to 53k pps, the signal amplitude first decreased abruptly from channel 2750 and then stabilized around channel 1260, whereas the time response was stable all along the irradiation (Fig. 12). The energy resolution was severely affected at this rate, a relative decrease of 8 % was observed after the irradiation.

The irradiation studies showed that rates higher than 53k pps had an impact on the energy response of a 10x10 mm$^2$ solar cell. Such behavior was observed for all cells sizes irradiated with rates above a hundred thousand pps. Nevertheless, the time response was only affected after irradiating with intensities above one million pps.

After the irradiation, the cell would continue to provide a signal amplitude 4 times lower (ch 800) in the energy spectrum. Interestingly, the time response was essentially not affected (Fig. 13).

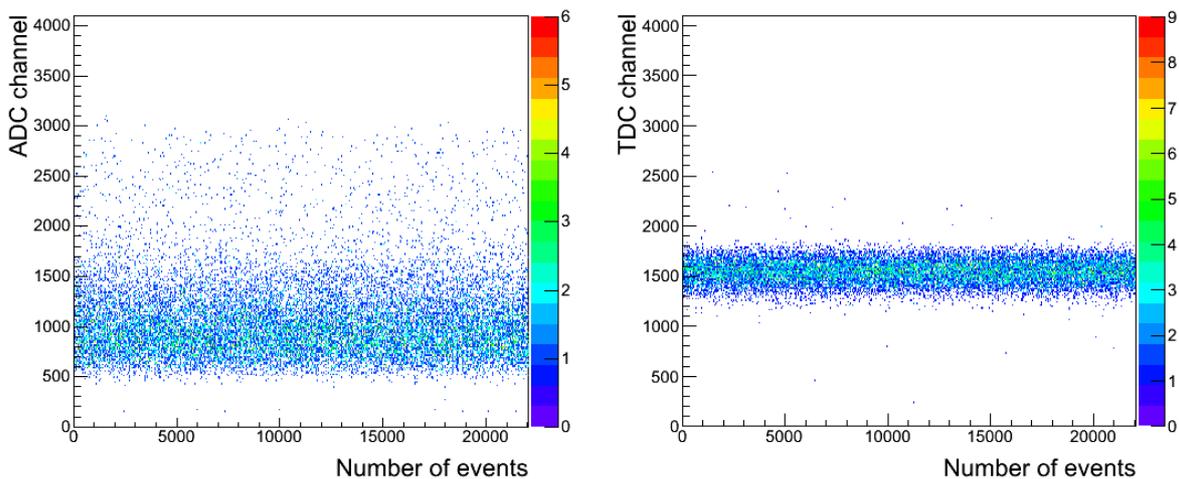

Figure 13- Energy and time response of a 10x10 mm$^2$ cell irradiated with $^{84}$Kr beam at 7 AMeV (225 pps) after having been irradiated with a total of 3427 million $^{84}$Kr ions with rates as high as 1 million pps.

We have also investigated an intermediate intensity range using a 10x10 mm$^2$ cell that was irradiated with a 3.8 AMeV $^{238}$U beam at a 4000 pps rate (Fig. 14). The experimental conditions were worse than for the Kr and Xe beams, which explains the observed tail spreading to larger energies, but the figure shows the stability of the energy response at higher rates and deposited energy.

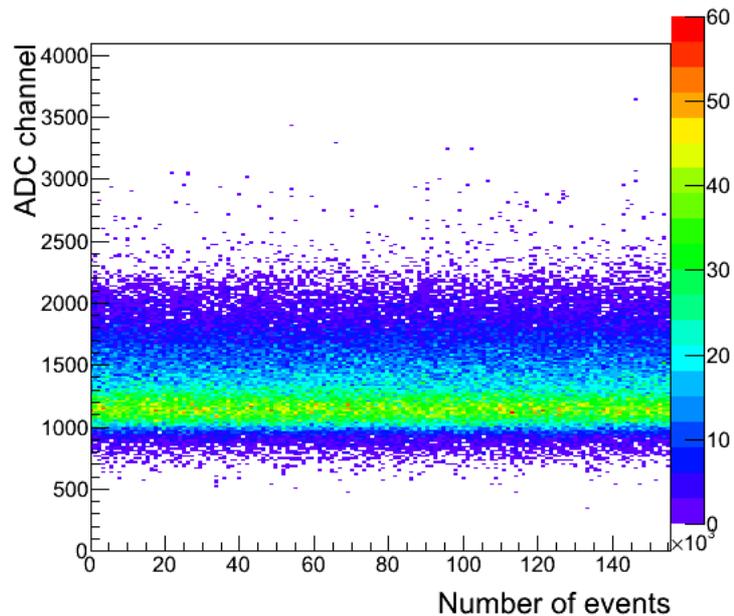

Figure 14 - Energy response of a germanium 10x10 mm$^2$ cell at a rate of 4000 pps with $^{238}$U beam at 3.8 AMeV.

## 6. Conclusions and Outlook

Solar cells were for the first time exposed to heavy ions of energies above 1 AMeV, at the GANIL facility, Caen, France. For such tests a new set-up was developed. This set-up was prepared considering very stringent possible vacuum requirements. All cells tested have responded to the heavy ion beams used and all energies, however the best results for energy and time resolution were observed for smaller cells of 10x10 mm$^2$. The energy resolution of a 10x10 mm$^2$ solar cell ranges from 1 to 3% while the time resolution ranges from 3.6 to 7 ns. These results are to some extent comparable to the ones obtained from a silicon detector with an energy resolution of about 0.5% and a time resolution between 1 and 4 ns. Regarding the behavior as a function of the beam intensity, we observe a stable behavior for rates ranging from few hundreds to few thousands pps and a clear loss of energy resolution and amplitude when irradiating a solar cell with $^{84}$Kr at 7 AMeV at a rate of 50 thousand pps. The time response was stable at all the rates.

The general behavior observed demonstrates that solar cells can be used to count heavy ions and measure time coincidences over a broad range of incident energies well above 1 A MeV and for rates as high as several thousand pps.

All the results obtained in these first exploratory measurements showed evidence of a promising heavy ion detector to be used for beam diagnostic or as heavy-ion detector in experiments with radioactive ion beams and storage rings. In the near future, we foresee additional studies to further investigate the behavior with beam energy, with different ions at similar energies and to compare the radiation resistance between the solar cells and a silicon detector. In particular, we aim at evaluating fluences and pulse-heights for a long and continuous irradiation. We also aim to perform further irradiations improving the beam diagnostics conditions to study the position sensitivity of the cells. Other improvements will be carried out regarding the pre-amplifiers, mainly to optimize their signal-to-noise ratio.

**Acknowledgements**

This work has been supported by the French défi interdisciplinaire NEEDS and by the European commission within the EURATOM FP7 Framework Program through the CHANDA (project no. 605203). This project as received funding from the European Union's Horizon 2020 research and innovation program under the Marie Sklodowska-Curie grant agreement No 834308 "NURING". B. Jurado gratefully acknowledges the support provided by the ExtreMe Matter Institute EMMI at the GSI Helmholtzzentrum fuer Schwerionenforschung, Darmstadt, Germany. We thank L. Hirsch for making possible the measurement of the solar cell impedance at the IMS laboratory of the University of Bordeaux. We thank the technical staff members of the AIFIRA facility. The AIFIRA facility is financially supported by the CNRS, the University of Bordeaux and the Région Nouvelle Aquitaine. Yu. A. Litvinov, J. Glorius and L. Varga acknowledge the funding from the European Research Council (ERC) under the European Union's Horizon 2020 research and innovation programme (grant agreement No 682841 "ASTRUm").

**References**

[1] G. Siegert et al., Nucl. Instrum. Methods Phys. Res. 164 (3) (1979) 437-438.

[2] N. Ajitanand et al., Nucl. Instrum. Methods Phys. Res. A 300 (2) (1991) 354-356.

[3] E. Liatard et al., Nucl. Instrum. Methods Phys. Res. A 267 (1988) 231-234.

[4] G. Kessedjian et al., Physics Letters B 692 (5) (2010) 297-301.

[5] G. Kessedjian et al., Phys. Rev. C 91 (2015) 044607.

[6] J. Koglin et al., Nucl. Instrum. Methods Phys. Res. A 854 (2017) 64-69.

[7] R. Pérez Sánchez et al., Nucl. Instrum. Methods Phys. Res. A 933 (2019) 63-70.

[8] T. Kirchartz et al., Phys. Chem. Chem. Phys. 17 (2015) 4007-4014.

[9] C. Hsieh et al., Electron Device Letters, IEEE 2 (1981) 103-105.

[10] F. B. McLean et al., IEEE Transactions on Nuclear Science 29 (1982) 2017-2023.

[11] G. C. Messenger et al., IEEE Transactions on Nuclear Science 29 (6) (1982) 2024-2031.

[12] B. Jurado et al., European Phys. J. Web of Conf. 146 (2017) 11006.

[13] Solar made, https://www.solarmade.com/, accessed: 2019-10-08.

[14] Space Azur solar power gmbh, http://www.azurspace.com/index.php/en/, accessed: 2019-10-08.

[15] S. Sorieul et al., Nucl. Instrum. Methods Phys. Res. B 332 (2014) 68–73.

[16] C. Theisen et al. Proceedings of the Second International Workshop on Nuclear Fission and Fission-Product Spectroscopy, The American Institute of Physics, Seyssins, France, (1998), pp. 143–150.

[17] G. F. Knoll, *Radiation, Detection and Measurement*, 3rd ed. (John Wiley & Sons, New York, 2000)

[18] H.W. Kraner et al., Nucl. Instrum. Methods Phys. Res. B 225 (1984) 615-618.

[19] V.A.J. Van Lint et al., Nucl. Instrum. Methods Phys. Res. A 253 (1987) 453-459.

[20] G. Hall et al., Nucl. Instrum. Methods Phys. Res. A 368 (1995) 199-214.

[21] G. Lindström et al., Nucl. Instrum. Methods Phys. Res. A 512 (2002) 30-43.

[22] F. Shiraishi et al., Nucl. Instrum. Methods Phys. Res. 69 (1969) 316-322.

[23] M. Kurokawa et al., IEEE Transactions on Nuclear Science 42 (3) (1995) 163-168.

[24] V. Eremin et al., JINST 13 P01019 (2018).

[25] C. Gautherin, PhD Thesis, CEA/Saclay DSM (1997), URL : https://inis.iaea.org/collection/NCLCollectionStore/_Public/31/017/31017434.pdf?r=1&r=1